\documentclass[sigconf, review=False, natbib=false]{acmart}

\renewcommand\footnotetextcopyrightpermission[1]{} 

\AtBeginDocument{%
  }

\setcopyright{rightsretained}  
\copyrightyear{2023}
\acmYear{2023}
\acmDOI{XXXXXXX.XXXXXXX}

\acmConference[ICAIF ’23]{}{November 27-29, 2023}{New York, NY, USA}

\acmPrice{}
\acmISBN{978-1-4503-XXXX-X/18/06}

\RequirePackage[
  datamodel=acmdatamodel,
  style=acmnumeric,
  ]{biblatex}

\begin{document}

\title{Multi-Factor Inception: What to Do with All of These Features?}

\author{Tom Liu}
\affiliation{%
  \institution{University of Oxford}
  \city{Oxford}
  \country{UK}
}


\author{Stefan Zohren}
\affiliation{%
  \institution{Oxford-Man Institute of Quantitative Finance, University of Oxford}
  \city{Oxford}
  \country{UK}
  }


\begin{abstract}
Cryptocurrency trading represents a nascent field of research, with growing adoption in industry. Aided by its decentralised nature, many metrics describing cryptocurrencies are accessible with a simple Google search and update frequently, usually at least on a daily basis. This presents a promising opportunity for data-driven systematic trading research, where limited historical data can be augmented with additional features, such as hashrate or Google Trends. However, one question naturally arises: how to effectively select and process these features? In this paper, we introduce Multi-Factor Inception Networks (MFIN), an end-to-end framework for systematic trading with multiple assets and factors. MFINs extend Deep Inception Networks (DIN) \cite{liu2023DIN} to operate in a multi-factor context. Similar to DINs, MFIN models automatically learn features from returns data and output position sizes that optimise portfolio Sharpe ratio. Compared to a range of rule-based momentum and reversion strategies, MFINs learn an uncorrelated, higher-$Sharpe$ strategy that is not captured by traditional, hand-crafted factors. In particular, MFIN models continue to achieve consistent returns over the most recent years (2022-2023), where traditional strategies and the wider cryptocurrency market have underperformed.

\end{abstract}





\maketitle

\section*{DISCLAIMER}
This work reflects the analysis and personal view of the author Tom Liu. No reader should interpret this work to represent the views of any third party. Assumptions, opinions, views and estimates constitute the author’s judgement as of the date given and are subject to change without notice and without duty to update. 


\section{Introduction}

Cryptocurrencies have emerged as a novel and increasingly popular asset class in recent years. Since the inception of Bitcoin \cite{nakamoto2008peer} in 2009, the global cryptocurrency market has grown rapidly, reaching a market capitalisation of \$1.2 trillion in 2023 \cite{CoinMarketCap}. This is fast approaching established equity indices, such as FTSE 100 and EURO STOXX 50. In a similar trend, researchers are actively exploring strategies to capitalise on cryptocurrency trading opportunities. 

Cryptocurrency markets exhibit parallels with equities, including high volatility, positive asset correlations, and idiosyncratic risk. Consequently, one approach involves adapting rule-based strategies initially developed for equities into the cryptocurrency domain. Studies have demonstrated the presence of analogous effects, such as momentum \cite{caporale2020momentum,li2021MAXmomentum,liu2022common,}, reversion \cite{leung2019constructing}, value and carry \cite{hubrich2017know}, in cryptocurrency data. Notably, \cite{liu2021risks} reveals a strong Time Series Momentum (TSMOM) effect in both price data and alternative data sources like Google searches and tweets. This indicates the wider applicability of technical indicator-based strategies beyond conventional price and volume data for cryptocurrencies.

In addition to traditional strategies, researchers increasingly apply Machine Learning (ML) techniques in cryptocurrency analysis. Although cryptocurrencies have a relatively short history, their decentralised nature facilitates access to higher frequency data, including minute-by-minute prices from platforms like Binance \cite{BinanceKaggle}, as well as alternative data sources \cite{BitInfoCharts, Blockchair}, which can help train more complex ML models. Transfer learning from related assets, such as Foreign Exchange (FX), has also been used to train Fused Encoder Networks in data-scarce scenarios \cite{poh2022transfer}.

Prior ML research often relies on hand-crafted input features. For example, \cite{poh2022transfer} extracts raw and normalised price returns over different time frames. While recent models can learn features directly from data, they are typically limited to a single input factor \cite{jiang2017cryptocurrency,liu2023DIN,lucarelli2020deep, zhang2021universalE2E,}. For ML models, we define factors as a type of input data, including price, volume or alternative data, such as hashrate. Moving to a multi-factor context with alternative data, \cite{qing2022fundamental} provides indicators for change, momentum, and decay in each factor as the inputs for their AutoEncoder model. This scenario presents a common challenge: when developing new multi-factor ML strategies, a crucial question is how to select and process features. This can significantly increase the dimensionality of the input for time series models that leverage prior historical data. Simultaneously, it restricts model inputs to a predefined set of hand-crafted features.

Our approach addresses these challenges by providing a single time series of returns for each asset and factor. We then develop a model that can learn useful features automatically from the data. To achieve this, we introduce Multi-Factor Inception Networks (MFIN) as a novel way to combine price and alternative data across multiple assets. This draws on prior work in Deep Inception Networks (DIN) \cite{liu2023DIN}, and extends the concept to multiple factors, including price, volume, hashrate and tweets. Our experiments demonstrate that MFIN strategies remain profitable after transaction costs and learn uncorrelated behaviours to traditional strategies, such as momentum and reversion. MFIN can also be combined with an existing portfolio of traditional strategies, contributing positively to Sharpe ratio and breakeven transaction costs, whilst limiting drawdowns.

\section{Data}

We evaluate strategy performance on 7 cryptocurrencies, according to price and volume data from CoinMarketCap (CMC) \cite{CoinMarketCap} over January 2018 to March 2023. These are spot rates against USD for BCH, BTC, DASH, DOGE, ETH, LTC, ZEC. Given that these coins are highly liquid, prior work has shown that CMC prices are reliable and accurately reflect the underlying market processes \cite{vidal2022cryptocurrency}.

To avoid survivorship bias, we use Wayback Machine \cite{WaybackMachine} to select the top 25 coins by market capitalisation on CMC as of January 2019: the start of the first test set. Stablecoins, such as Tether and USD coin, are excluded. We then filter to the final 7 coins according to the availability of alternative data on BitInfoCharts (BIC) \cite{BitInfoCharts} in January 2019. We supplement and cross-check alternative data with Blockchair (BC) \cite{Blockchair}, where data for 6 coins is available from March 2019. The final coin (ZEC) is added to BC in 2020.

Google Trends data $G$ is missing for 3 of 7 coins in BIC, so we obtain this from the Google Trends website \cite{GoogleTrends}. Due to website limits, we download data in 90-day segments to allow for daily resolution. We link segments with backwards proportional adjustment \cite{stridsman1998data} from Eq. \ref{eqn:rad_price}. This preserves percentage changes, since Google Trends are scaled in proportion to the largest signal in each segment.
\begin{gather}
        G_{adjusted} = G_{raw} * \frac{g_2}{g_1},  \label{eqn:rad_price} \\
        g_1 = \text{last datum of preceding segment on roll date}, \nonumber\\
        g_2 = \text{first datum of new segment on roll date}. \nonumber 
\end{gather}

To avoid lookahead bias, all strategies trade at market open on date $t$, using price and alternative data from prior dates up to $t-1$. Although new open prices are available at date $t$, we do not include this to improve tradeability. The features from each data source are listed in Table \ref{table:Data_Sources}. 

\begin{table}[h]
\centering
\caption{Features from CoinMarketCap (CMC), BitInfoCharts (BIC) and BlockChair (BC) data sources. All monetary values are in USD.}
\begin{tabular}{@{\extracolsep{4pt}} c | c | c} 
 \toprule
 \textbf{CMC} & \textbf{BIC} & \textbf{BC} \\
 Price & Alternative & Alternative \\
 \midrule\midrule
 open           & transactions & fee-reward ratio \\
 high           & block size &  chain size increase \\
 low            & sent addresses &  coin days destroyed \\
 close          & sent USD &  cost per transaction \\
 volume         & difficulty &   \\
 market cap.    & mining profitability &   \\
                & hashrate &   \\
                & av. transaction size&   \\
                & av. transaction value&   \\
                & confirmation time &   \\
                & tweets &   \\
                & google trends &   \\
\bottomrule
\end{tabular}
\label{table:Data_Sources}
\end{table}

\section{Traditional Strategies}
We first apply rule-based strategies that capture momentum and reversion behaviour separately.

\subsection{Momentum Strategies}

For each alternative feature $j$ from the BIC and BC data sources, we consider a Time Series Momentum (TSMOM) strategy that buys/shorts assets with positive/negative signal strength,
\begin{equation}
\label{eq:TS_weights}
w_{i,t} = \mathrm{sign}(s_{i,t}).
\end{equation}


\subsubsection{Moskowitz, Ooi \& Pedersen (MOP) \cite{moskowitz2012time}}

MOP is the seminal TSMOM strategy. The authors propose a score based on past 12-month returns for a strategy with monthly investment horizon. In this paper, we modify the signal $s_{i,t} = r_{i,j,t}^{(k)}$ to be returns over past $k \in \{5, 21, 63, 126, 252\}$ days.


\subsubsection{Baz et al. (BAZ) \cite{baz2015dissecting}}

BAZ proposes using MACD indicators to capture the cross-over of exponentially-weighted moving averages (EWMA). When the EWMA with short timescale $S_k \in \{ 4, 8, 16, 32 \}$ is greater than the EWMA with long timescale $L_k \in \{ 12, 24, 48, 96 \}$, then we expect the asset to have upwards momentum and vice versa. The resulting BAZ signal is $s_{i,t} = \mathrm{MACD}_{i,j,t}^{(k)}$.

\subsection{Reversion Strategies}


\subsubsection{Bollinger Bands (REV) \cite{bollinger2002bollinger}}

REV is a reversion strategy, based on the spread $\delta_{i,j,t} = r_{i,open,t}^{(k)} - r_{i,j,t}^{(k)}$ between the open price and an alternative feature $j$ on the same date $t$. The return length $k \in \{1, 5, 10, 21\}$ days is a tuneable parameter. We compute a z-score signal $s_{i,t}$ from $\delta_{i,j,t}$ in a rolling 63-day exponentially weighted window. The portfolio weights short each asset according to the entry $z_u \in \{1.5, 1.75, 2.0\}$ and exit $z_l \in \{0.5, 0.75, 1.0\}$ z-score thresholds.
\begin{equation}
    w_{i,t} = 
    \begin{cases}
      -\mathrm{sign}(s_{i,t}), & \text{if } |s_{i,t}| \geq z_u \text{ and } w_{i,t-1} = 0\\
      w_{i,t-1}, & \text{if } |s_{i,t}| \geq z_l \text{ and } w_{i,t-1} \neq 0\\
      0, & \text{otherwise}.\\
    \end{cases} 
\end{equation}

We verify that the spread $\delta_{i,j,t}$ is stationary by applying an Augmented Dicky Fuller (ADF) test, requiring a p-value $\leq 1\%$.

\subsection{Additional Benchmarks}


\subsubsection{Combined (CMB)}

CMB is a volatility-scaled portfolio of MOP, BAZ and REV strategies and we compare this with the inherent ability of MFIN models to combine information from different input factors.


\subsubsection{Long-only}

This benchmark represents market returns and assigns equal weight $w_{i,t} = 1$ to all assets $i$.

\subsection{Portfolio Construction}

Each strategy outputs portfolio weights $w_{i,t}$ for all assets $i$ at time $t$. We perform volatility scaling across $N_A=7$ assets to $\sigma_{tgt}=15\%$, to improve Sharpe ratio and decrease the likelihood of extreme returns \cite{harvey2018impactofvolscaling}. The portfolio returns $R_{p,t}$ are defined as,
\begin{equation}
\label{eqn:portfolio_returns}
  R_{p,t+1} = \frac{\sigma_{tgt}}{N_A}\sum_{i=1}^{N_A} \frac{w_{i,t}}{\sigma_{i,t}}\cdot r_{i,t+1} - C \left |\frac{w_{i,t}}{\sigma_{i,t}} - \frac{w_{i,t-1}}{\sigma_{i,t-1}} \right |,
\end{equation}
given asset returns $r_{i,t}$, annualised ex-ante volatility $\sigma_{i,t}$ and transaction cost coefficient $C$. This is consistent with \cite{lim2019enhancing,liu2023DIN,wood2021trading,wood2022slow}. 

When comparing different strategies, we apply a second layer of volatility scaling to portfolio returns, so that each strategy takes a similar overall risk. During live trading, it is not possible to scale portfolio volatility to exactly $\sigma_{tgt}=15\%$, since future volatility can only be estimated. For more realistic results, we perform the second stage of volatility scaling with a rolling 21-day exponentially-weighted standard deviation. This is applied to all strategies, including Long-only.

As shown in \cite{lim2019enhancing,liu2023DIN}, the performance of Machine Learning models, such as LSTM and DIN, varies across random initialisations. Therefore, to ensure robust results, we ensemble MFINs over 10 random seeds. Each model is trained with the same hyperparameters and we take a simple average of their outputs.

\subsection{Back-testing Details}
\label{section:backtesting}
We optimise MOP, BAZ, and REV strategies across features $j$ and parameters $k$, $z_u$, $z_l$. For data exploration purposes, we produce an overestimate of strategy performance by selecting the two best feature-parameter combinations, according to Sharpe ratio. This is an ex-post analysis, using all available test data from April 2019.

\begin{figure}[h]
  \centering
  \includegraphics[width=\linewidth]{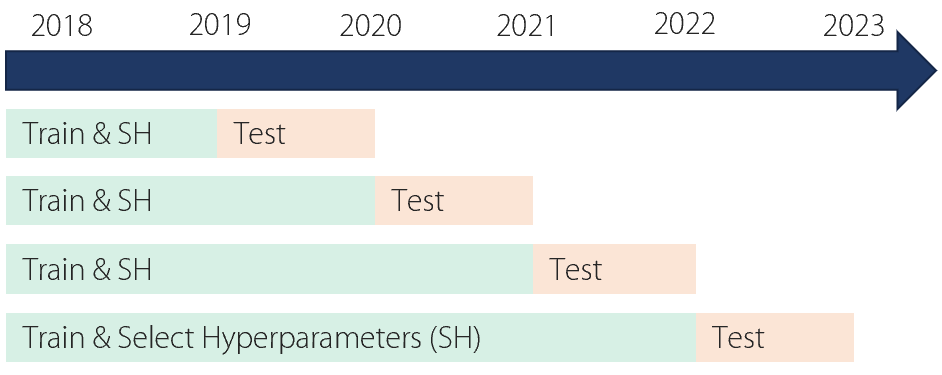}
  \caption{Expanding-window train-test splits.}
  \label{fig:TrainTestSplits}
\end{figure}

However, realistic strategies must avoid lookahead bias. We simulate this by using expanding-window train-test splits. This is the same procedure for MFIN models, and is shown in Figure \ref{fig:TrainTestSplits}. We rank all feature-parameter combinations based on Sharpe ratio in each training set, and evaluate an equally weighted portfolio of the top two combinations in the corresponding test set. For diversification, we ensure that the chosen combinations do not share the same feature. This prevents selecting similar strategies with slightly different signal parameters.

\section{Multi-Factor Inception Networks}
\label{section:Methods_MFIN}

\begin{figure*}[!thbp]
  \includegraphics[width=0.99\textwidth]{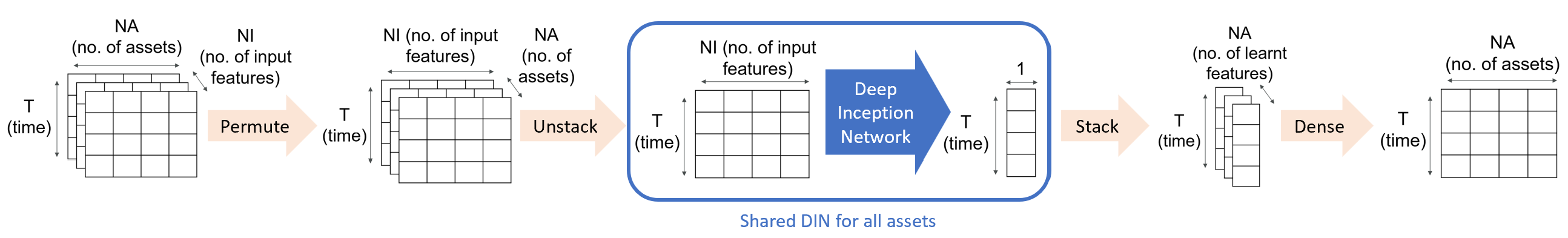}
  \caption{High-level model architecture for Multi-Factor Inception Networks (MFINs).}
  \label{fig:ModelArchitecture_HighLevel}
\end{figure*}

\begin{figure*}[!thbp]
  \includegraphics[width=0.99\textwidth]{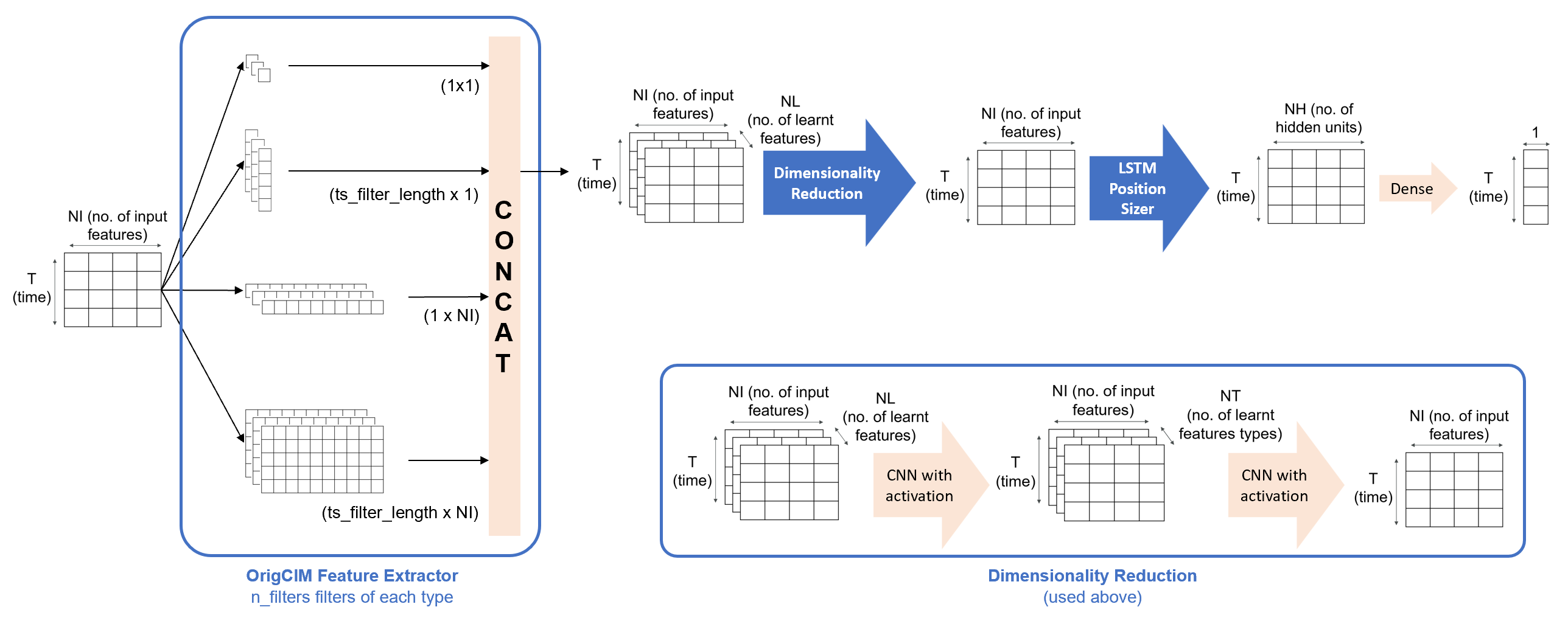}
  \caption{OrigCIM-LSTM variant of Deep Inception Network (DIN).}
  \label{fig:ModelArchitecture_OrigCIM-LSTM}
\end{figure*}

Deep Inception Networks (DINs) \cite{liu2023DIN} are an end-to-end framework for systematic investing, which learn features and position sizing directly from daily returns. Although originally developed for single-factor scenarios, we can extend DINs to multiple input features with Multi-Factor Inception Networks (MFIN). As shown in Figure \ref{fig:ModelArchitecture_HighLevel}, we process a ($T \times N_I$) matrix of standardised daily returns for each asset with a DIN and add a dense output layer to optimise the multi-asset portfolio. $T$ is the sequence length and $N_I$ is the number of input features. The same DIN is used for all assets, which reduces the model complexity in proportion to the total available data. For example, increasing the number of assets $N_A$ from 1 to 50 would not increase the number of model parameters by $50 \times$.

Following the underlying DINs, the MFIN model outputs portfolio weights that directly optimise Sharpe ratio, adjusted for transaction costs and correlation to Long-only. 
\begin{gather}
\label{eqn:loss}
    \mathrm{Loss} = -\sqrt{252} \cdot\frac{\mathrm{Mean}(R_{p,t})}{\mathrm{Std}(R_{p,t})} + K\cdot|\rho|,
\end{gather}
where $R_{p,t}$ are volatility-scaled portfolio returns adjusted for transaction costs. $R_{b,t}$ are portfolio returns of the Long-only benchmark and $\rho$ is Pearson correlation between $R_{p,t}$ and $R_{b,t}$. 

As defined in Eq. \ref{eqn:portfolio_returns}, $R_{p,t}$ depends on transaction cost coefficient $C$. A large value of $C$ encourages lower portfolio turnover. The correlation regularisation coefficient $K$ helps prevent MFIN models from adopting a static behaviour that may fail to generalise out-of-sample. Hyperparameters $C$ and $K$ are automatically tuned based on losses in the validation sets. In this project, we set $K_{valid} = C_{valid} = 0$ to avoid manually scaling the contribution of each term in the validation loss.

\subsection{Model Inputs and Outputs}
\label{section:Architecture_IO}

MFINs take 2 input tensors, of size ($T$ x $N_A$ x $N_I$) and ($T$ x $N_A$ x $2$), respectively. Here, we use sequence length $T = 100$, number of assets $N_A = 7$ in the portfolio, number of input features $N_I = 22$, from both price and alternative data.
\begin{enumerate}
    \item \textbf{Past returns $X_t^{(j)}$}: daily returns for each asset $i$ and input feature $j$, standardised by 63-day exponentially-weighted standard deviation $\frac{r_{i,j,t}}{\sigma{i,j,t}}$. The last row of $X_t^{(j)}$ is the most recent returns data, computed from $t-2$ to $t-1$ to prevent lookahead bias.
    \item \textbf{Future returns $Y_t^{(1)}$}: $X_t^{(open)}$ shifted forwards by 2 days and scaled to $\sigma_{tgt}$. The last row is ``one-step-ahead" returns in open price, from $t$ to $t+1$.
    \item \textbf{Volatility scaling $Y_t^{(2)}$}: volatility scaling factors $\frac{\sigma_{tgt}}{\sigma_{i,1,t}}$, corresponding to $Y_t^{(1)}$.
\end{enumerate}
At each timestep $t$, DIN models predict a vector of position sizes $w_{i,t} \in [-1,1]$ for portfolio construction. The $X_t$ tensor is used for predictions and $Y_t$ is used to evaluate the loss function. In this paper, we use a batch size equivalent to sequence length $T$.

\subsection{Deep Inception with OrigCIM-LSTM}

Due to the limited amount of cryptocurrency data, we favour simpler Deep Inception Network (DIN) models. For a fixed $N_I = 22$, one potential selection is OrigCIM-LSTM. The architecture of this DIN variant is illustrated in Figure \ref{fig:ModelArchitecture_OrigCIM-LSTM}. As shown in \cite{liu2023DIN}, OrigCIM has fewer trainable parameters than more scalable Feature Extractors, such as FlexCIM, when the input width is relatively small. However, in order to capture cross-factor interactions between input features using the Feature Extractor, we do not use simpler Time Series models, such as DeepLOB \cite{zhang2019deeplob}.


\subsubsection{Original Custom Inception Module (OrigCIM)}

OrigCIM is a Feature Extractor that uses a single layer of convolutional filters. We extract features in each time series (TS), across input features (CS), and across both time and input features (combined). These correspond to filters of size ($ts\_filter\_length \times 1$), ($1 \times N_I$), and ($ts\_filter\_length \times N_I$), respectively. ($1 \times 1$) filters propagate the original returns data to the next stage, with a learnt scaling factor. In total, there are $N_T = 4$ feature types, which are learnt in parallel with an Inception Module structure. We use $n\_filters$ filters of each type to learn a variety of TS, CS and combined features. This intermediate feature tensor has size ($T \times N_A \times N_F$), where the number of learnt features $N_F = N_T \times n\_filters$. Finally, we perform Dimensionality Reduction by convolving across learnt features.


\subsubsection{Long Short-Term Memory (LSTM) \cite{lim2019enhancing, hochreiter1997LSTM}}
\label{section:Architecture_LSTM}

To constrain model complexity, the largest choice of hyperparameter $ts\_filter\_length$ is 20. Therefore, we use an LSTM in the Position Sizer to capture longer-term interactions in the data. For $N_H = hidden\_layer\_size$ LSTM cells, the output size is ($T \times N_H$). Different to the original DINs proposed in \cite{liu2023DIN}, a final dense layer in the OrigCIM-LSTM variant produces an output width of 1. Portfolio optimisation across assets occurs within the encapsulating MFIN model instead.

\subsection{Model Training}
Following \cite{lim2019enhancing,poh2021building,wood2021trading,wood2022slow}, we train our model with an expanding window approach with a 90\%/10\% train/validation split for hyperparameter optimisation. The cryptocurrency dataset uses 1-year increments, as shown in Figure \ref{fig:TrainTestSplits}. 

All MFIN models are implemented in \textit{Tensorflow}. We customise the model training pipeline so that training losses are calculated from ``one-step-ahead" predictions of the entire batch, and disable data shuffling during training. This ensures that the transaction costs in the loss function of Eq. \ref{eqn:loss} are correct. Consistent with \cite{lim2019enhancing,liu2023DIN,poh2021building,wood2021trading,wood2022slow,}, we use the Adam optimiser \cite{kingma2014adam}. 

Given prior work which shows HB to be effective for tuning the underlying DINs \cite{liu2023DIN}, we use 30 iterations of Hyperband (HB) for hyperparameter optimisation. MFIN models train for a maximum of 250 epochs. Early stopping after 25 epochs of non-decreasing validation loss is used to avoid overfitting to training data. A full list of fixed and tuneable parameters is provided in Table \ref{table:hyperparams}.

\begin{table}[!ht]
\centering
\caption{\centering{Hyperparameters for MFIN models.}}
\begin{tabular}{ @{\extracolsep{4pt}} c| c | c } 
 \toprule
 Parameter & Category & Value \\
 \midrule\midrule
 Objective & Fixed & Eq. \ref{eqn:loss} \\[0.5ex] 
 Max epochs  & Fixed & 250 \\[0.5ex] 
 Early stopping  & Fixed & 25 \\[0.5ex] 
 Train/valid ratio & Fixed & 90\%/10\% \\[0.5ex] 
 Batch size & Fixed & 100 \\[0.5ex]
 \midrule
 Sequence length $T$ & Fixed & 100 \\[0.5ex] 
 Activation  & Fixed & ELU \\[0.5ex] 
 $C_{valid}$  & Fixed & 0  \\ [0.5ex] 
 $K_{valid}$  & Fixed & 0   \\[0.5ex] 
 \midrule
 $C$  & Tuned & [0, 0.5, 1, 2, 5]    \\[0.5ex] 
 $K$  & Tuned & [0, 1, 2, 5]    \\[0.5ex] 
 $dropout\_rate$  & Tuned & [0.1, 0.2, 0.3]   \\[0.5ex] 
 $learning\_rate$ & Tuned & [1e-3, 1e-4, 1e-5] \\[0.5ex]
 hidden\_layer\_size, $N_H$ & Tuned & [32, 64, 96, 128] \\[0.5ex] 
 n\_filters & Tuned & [16, 32, 48, 64] \\ [0.5ex] 
 ts\_filter\_length & Tuned & [3, 5, 10, 15, 20] \\[0.5ex]
 \midrule
 max\_epochs & Hyperband & 10 \\ [0.5ex] 
 hyperband\_iterations & Hyperband & 1  \\[0.5ex] 
 factor & Hyperband & 3  \\[0.5ex] 
 \bottomrule
\end{tabular}
\label{table:hyperparams}
\end{table}

\section{Results and Discussion}

\subsection{Metrics}
We evaluate strategies using similar metrics to \cite{liu2023DIN}, with modifications due to dynamic volatility scaling at the portfolio level.
\begin{enumerate}
    \item \textbf{Profitability}: mean annual returns ($MAR$).
    \item \textbf{Risk}: annualised volatility ($VOL$) and maximum drawdown ($MDD$). Since strategies can have differences in $VOL$, it is most meaningful to compare $MDD$ as a number of standard deviations $\sigma$ (i.e. a multiplier on $VOL$).
    \item \textbf{Risk-adjusted profitability}: annualised $Sharpe$, $Sortino$, and $Calmar$ ratios. 
    \item \textbf{Correlation ($CORR$)}: low (absolute) Pearson rank correlation coefficient to Long-only helps produce consistent returns and avoid drawdowns during market crashes.
    \item \textbf{Transaction costs}: breakeven transaction cost ($BRK$) is expressed in basis points (bps), after volatility scaling at the asset level and portfolio level with $\sigma_{i,t}$ and $\sigma_{t}$, respectively.
    \begin{gather}
        BRK = \frac{\sum_{t=1}^{T}\sum_{i=1}^{N_A} \frac{w_{i,t}}{\sigma_{i,t}\cdot\sigma_{t}}\cdot r_{i,t+1}}{\sum_{t=1}^{T}\sum_{i=1}^{N_A}|\frac{w_{i,t}}{\sigma_{i,t}\cdot\sigma_{t}} - \frac{w_{i,t-1}}{\sigma_{i,t-1}\cdot\sigma_{t-1}}|}
    \end{gather}
    \item \textbf{Statistical significance}: we interpret the Probabilistic Sharpe Ratio ($PSR$) \cite{bailey2012sharpe} as significant when above a 99\% confidence level against a benchmark $Sharpe$ of 0. Minimum Track Record ($MTR$) is the minimum number of observations, in days, needed for a significant result.
\end{enumerate}

\subsection{Performance of Traditional Strategies}

\begin{table*}[!tb]
\centering
\caption{\centering{
    Data exploration for traditional strategies. These strategies cannot be implemented with only ex-ante information. Best values in comparable metrics are \underline{underlined}.}}
\begin{tabular}{@{\extracolsep{4pt}}lll|ccc|cccc}
\toprule 
 Strategy  & Feature & Signal Parameters & Sharpe & Sortino & Calmar & VOL & MDD & CORR & BRK  \\[0.5ex]
 &  & & &  &  & \% & $\sigma$ & \% & bps  \\[0.5ex]
  \midrule\midrule
 MOP & sent addresses & $k = 21$ & 1.67  & 2.62 & 1.82 & 14.0 & 0.98  & 17.9 & 28.3 \\[0.5ex] 
 MOP & hashrate  & $k = 21$ & 1.49  & 2.27 & 1.39 & 14.0 & 1.14 & 48.6  & 29.9 \\[0.5ex] 
\midrule
 BAZ & sent addresses  & $(S_k,L_k) = (4,12)$ & 1.52  & 2.37 & 1.08 & 14.0 & 1.49 & -15.4 & 47.7  \\[0.5ex] 
 BAZ & block size & $(S_k,L_k) = (32,96)$ & 1.47  & 2.24 & 1.31 & 14.4  & 1.18 & \underline{-0.2} & \underline{193.7} \\[0.5ex] 
\midrule
 REV & coin days destroyed  & $(k,z_u,z_l) = (5,1.75,0.75)$ & 1.64 & 2.65 & 1.43 & 13.7 & 1.23 & 83.4 & 35.7 \\[0.5ex] 
 REV & google trends  & $(k,z_u,z_l) = (5,1.75,0.75)$ & 1.52 & 2.80  & 1.56 & 11.5 & 1.01 & 68.5 & 63.5 \\[0.5ex] 
 \midrule
 CMB & - & - & \underline{3.07} & \underline{5.40} & \underline{4.50} & 14.0 & \underline{0.83} & 37.7 & 54.0  \\[0.5ex] 
 \midrule
 Long-only & - & - & 1.03 & 1.47 & 0.66 & 14.4 & 1.56 & - & -  \\[0.5ex]  
 \bottomrule
\end{tabular}
\label{table:Results_DataExploration}
\end{table*}

\begin{table*}[!tb]
\centering
\caption{\centering{
     Comparison of MFIN performance against realistic traditional strategies. Best values in comparable metrics are \underline{underlined}. Significant PSR, compared to benchmark Sharpe ratio of 0, are highlighted in \textbf{bold}.}}
\begin{tabular}{@{\extracolsep{4pt}}l|ccc|ccc|cccc|cc}
\toprule 
 Strategy  & MAR & HR & PNL & Sharpe & Sortino & Calmar & VOL & MDD & CORR & BRK & PSR & MTR \\[0.5ex]
    & \% & \% &  &  &  &  & \% & $\sigma$ & \% & bps & \% & days \\[0.5ex]
\midrule\midrule
 MFIN & 24.2 & \underline{54.5} & 1.08 & 1.64 & 2.55 & 2.05 & 13.8 & 0.86 & -8.2 & 17.5 & \textbf{99.9} & 727 \\[0.5ex] 
\midrule
 MOP & 9.7 & 51.1 & 1.07 & 0.74 & 1.11 & 0.75 & 13.7 & 0.95 & \underline{0.3} & 9.1 & 93.0 & 3539 \\[0.5ex] 
 BAZ & 17.9 & 52.1 & 1.11 & 1.25 & 1.89 & 0.91 & 13.9 & 1.41 & -7.2 & \underline{36.8} & \textbf{99.3} & 1259 \\[0.5ex] 
 REV & 11.1 & 52.7 & 1.06 & 0.85 & 1.28 & 0.48 & 13.4 & 1.72 & 81.6 & 27.6 & 95.4 & 2724 \\[0.5ex] 
 \midrule
 CMB & 21.0 & 52.0 & \underline{1.15} & 1.42 & 2.21 & 1.01 & 14.1 & 1.48 & 25.0 & 20.4 & \textbf{99.8} & 963 \\[0.5ex] 
 CMB + MFIN & \underline{29.1} & 54.1 & 1.14 & \underline{1.89} & \underline{3.05} & \underline{3.21} & 14.0 & \underline{0.64} & 15.7 & 20.9 & \underline{\textbf{100.0}} & \underline{535} \\[0.5ex] 
 \midrule
 Long-only & 14.8 & 53.7 & 99.6 & 1.03 & 1.47 & 0.66 & 14.4 & 1.56 & - & - & 97.8 & 1898 \\[0.5ex] 
\bottomrule
\end{tabular}
\label{table:Results_RealisticStrategies}
\end{table*}


\subsubsection{Data Exploration}
Data exploration deliberately overestimates strategy performance, but is useful to check for momentum and reversion behaviour in the dataset.

In Table \ref{table:Results_DataExploration}, we list the top 2 feature-parameter combinations for each strategy type. We find that the number of unique sender addresses (\textit{sent addresses}), is a strong momentum indicator and provides the highest $Sharpe$ for both trend and crossover indicators. 

Both fast and slow momentum are present. This can be seen from the best BAZ strategies using the shortest and longest MACD indicators, respectively, without much difference in performance before transaction costs. Notably, BAZ with \textit{block size} has low correlation to Long-only despite a slow signal. This is because the strategy is highly correlated during the bull market of 2021, and anti-correlated through the drawdown thereafter.

Although the 6 selected strategies have similar Sharpe ratios between \textbf{1.4} to \textbf{1.7}, combining information from different features is still beneficial. A simple volatility-scaled CMB portfolio increases $Sharpe$ to \textbf{3.07}, with a high breakeven transaction cost of \textbf{54.0 bps} (basis points). This shows that alternative data can add value to cryptocurrency trading strategies. However, we expect performance of realistic strategies, even simple ones, to decay out-of-sample.

\subsubsection{Realistic Strategies}
\label{section:Results_RealisticStrategies}

An implementable strategy must instead make investment decisions based wholly on past data, without lookahead bias from knowing which feature-parameter combinations perform well on the entire test set. This is achieved through the train-test splits described in Section \ref{section:backtesting}.

We tabulate results for realistic MOP, BAZ, and REV strategies in Table \ref{table:Results_RealisticStrategies}. Compared to the Data Exploration strategies, performance for MOP and REV decays by approximately 50\%. Only BAZ retains a similar $Sharpe$: the \textit{sent addresses} MACD signal is consistently profitable throughout the test set and and can be readily identified based on only ex-ante data. The full list of feature-parameter combinations is provided in Appendix \ref{apdx:FeatureParameterCombinations}.

The volatility-scaled CMB portfolio still improves Sharpe, but this is less noticeable: a 13\% increase from the best underlying strategy, compared to the 83\% increase achieved during data exploration. As discussed in Sections \ref{section:Methods_MFIN} and \ref{section:Results_MFIN}, a fully data-driven approach to combining factors may be useful for extracting further value from alternative data.

\begin{figure}[!htb]
\centering
    \includegraphics[width=0.47\textwidth]{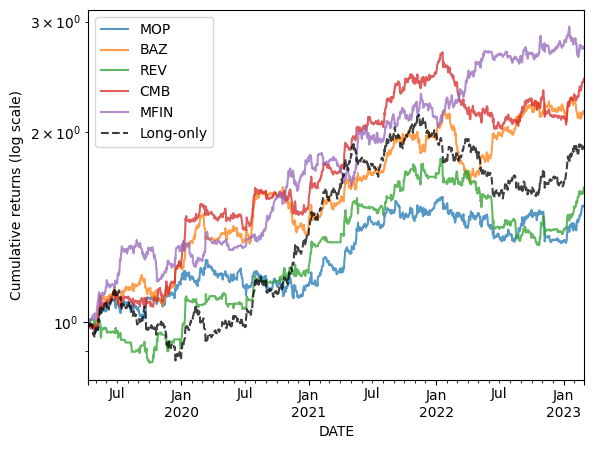}
    \caption{Cumulative returns for MFIN, traditional strategies without lookahead bias and Long-only at 0 transaction costs.}
    \label{fig:Strategy_returns}
\end{figure}

\subsection{Performance of MFIN Models}
\label{section:Results_MFIN}

\begin{table}[!hbtp]
\centering
\caption{\centering{
    Strategy performance at different transaction cost coefficients $C$. The best strategy at each $C$ is \underline{underlined}. Strategies that outperform a volatility-scaled Long-only benchmark are highlighted in \textbf{bold}.}}
\begin{tabular}{@{\extracolsep{4pt}} lcccccc}
\toprule   
{Strategy} & \multicolumn{6}{c}{Sharpe after $C$ (bps) of costs } \\[0.5ex]
 \cmidrule{2-7} 
  & 0.0 & 2.5 & 5.0 & 7.5 & 10.0 & 12.5 \\[0.5ex]
\midrule \midrule
 MFIN & \textbf{1.64} & \textbf{1.41} & \textbf{1.18} & 0.94 & 0.71 & 0.48 \\[0.5ex]
\midrule
 MOP & 0.74 & 0.54 & 0.34 & 0.14  & -0.06 & -0.26 \\[0.5ex]
 BAZ & \textbf{1.25} & \textbf{1.17} & \textbf{1.09} & 1.01 & 0.93 & 0.84 \\[0.5ex]
 REV & 0.85 & 0.77 & 0.70 & 0.63 & 0.55 & 0.48 \\[0.5ex]
 \midrule
 CMB & \textbf{1.42} & \textbf{1.25} & \textbf{1.08} & 0.90  & 0.73 & 0.56 \\[0.5ex]
 CMB + MFIN & \underline{\textbf{1.89}} & \underline{\textbf{1.67}} & \underline{\textbf{1.44}} & \underline{\textbf{1.21}} & 0.97 & 0.74 \\[0.5ex]
 \midrule
 Long-only & 1.03 & 1.03 & 1.03 & 1.03 & \underline{1.03} & \underline{1.03} \\[0.5ex]
\bottomrule
\end{tabular}
\label{table:Results_TransactionCosts}
\end{table}

\begin{figure}[!hbtp]
\centering
    \includegraphics[width=0.45\textwidth]{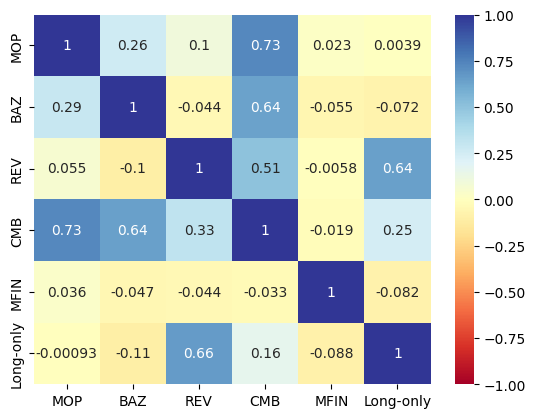}
    \caption{Correlation between MFIN, traditional and benchmark strategies. Values above and below the diagonal are Pearson and Spearman correlation coefficients, respectively.}
    \label{fig:Strategy_corr}
\end{figure}

We apply MFIN to standardised daily returns in each factor. This reduces the dimensionality of the input and features are instead learnt automatically from the data.

As shown in Table \ref{table:Results_RealisticStrategies}, MFIN outperforms traditional strategies and benchmarks prior to transaction costs. Inspecting cumulative returns in Figure \ref{fig:Strategy_returns}, MFIN performs consistently across time and is able to adapt to varying market conditions. For example, MFIN is the only strategy to avoid a large drawdown following November 2021, as investors sell risk assets including equities and cryptocurrencies. Accounting for returns, $Calmar = \textbf{2.05}$ is a threefold increase compared to Long-only, and double that of CMB.

Considering fees, MFIN is also able to deliver positive cost-adjusted Sharpe ratios up to $BRK = \textbf{17.5 bps}$, above the standard Binance trading fee of 10.0 bps \cite{BinanceFees}. At lower transaction costs up to 5.0 bps, MFIN maintains a higher $Sharpe$ than traditional strategies and benchmarks. Beyond 7.5 bps, BAZ, CMB and Long-only outperform due to less turnover, as shown in Table \ref{table:Results_TransactionCosts}. In particular, BAZ trades on slower MACD signals and REV holds no position until identifying a reversion opportunity. 

In Figure \ref{fig:Strategy_corr}, we plot a correlation matrix for traditional, MFIN and benchmark strategy returns. Pearson correlation coefficients are shown above the leading diagonal, and these agree with the Spearman coefficients below the diagonal. We find that MFIN is minimally correlated to all strategies. This shows that MFIN learns patterns that are fundamentally different from traditional rule-based strategies. We can also show that such ML strategies contribute additively to traditional strategies: a volatility-scaled combination of the CMB and MFIN portfolio weights (CMB + MFIN) has higher $Sharpe =$ \textbf{1.89} and $BRK =$ \textbf{20.9 bps}, whilst experiencing lower $MDD =$ \textbf{0.64 $\sigma$} than either underlying strategy.

\section{Conclusion}

We analyse the predictive power of alternative data for cryptocurrency trading. Momentum and reversion patterns are observed in the data, which can be combined additively to enhance individual Sharpe ratios from $\textbf{1.4-1.7}$ to a combined $Sharpe = \textbf{3.07}$. Removing lookahead bias, BAZ strategies with MACD indicators continue to perform out-of-sample, with only an 18\% decay in Sharpe ratio. 

We introduce Multi-Factor Inception Networks (MFIN) as a fully data-driven approach to learn from multiple factors, on multiple assets. MFIN has higher Sharpe ratio than traditional strategies and benchmarks up to $\textbf{5 bps}$, with a breakeven transaction cost of $BRK = \textbf{17.5 bps}$. For risk-sensitive investors, MFIN performs consistently over time and avoids large drawdowns, with a threefold increase in $Calmar = \textbf{2.05}$ compared to the underlying cryptocurrency market. This data-driven strategy can also be combined with traditional strategies to further improve performance to $Sharpe = \textbf{1.89}$, $Calmar = \textbf{3.21}$ and $BRK = \textbf{20.9 bps}$.

Future work can incorporate additional alternative data, such as Reddit subscribers and Github stars, which represent the popularity of a cryptocurrency on social media and within the technical community, respectively. As MFINs are a general framework for multi-factor models, another avenue of research could focus on enhancing the performance and interpretability of the underlying Deep Inception Network (DIN). One potential DIN variant is DeepLOB-TFT, which instead learns cross-factor interactions in the interpretable Temporal Fusion Transformer \cite{lim2021temporal, wood2021trading} Position Sizer. 

\begin{table*}[!ht]
\centering
\caption{\centering{Feature-parameter selections for rule-based strategies.}}
\begin{tabular}{ c | c c | c c | cc}
\toprule
 Test & \multicolumn{2}{c}{MOP} & \multicolumn{2}{c}{BAZ} & \multicolumn{2}{c}{REV}\\[0.5ex] 
set & Feature & Parameters & Feature & Parameters & Feature & Parameters  \\[0.5ex] 
\midrule \midrule   
2019 & transactions & $k=5$ & sent addresses & $(S_k,L_k) = (4,12)$ & sent USD & $(k,z_u,z_l) = (21,1.75,0.75)$  \\[0.5ex] 
- 2020 & sent addresses & $k=5$ & confirmation time & $(S_k,L_k) = (16,48)$ & google trends & $(k,z_u,z_l) = (5,1.5,1.00)$  \\[0.5ex] 
\midrule 
2020  & transactions & $k=5$ & confirmation time & $(S_k,L_k) = (32,96)$ & sent USD & $(k,z_u,z_l) = (21,1.75,0.75)$  \\[0.5ex] 
- 2021 & sent USD & $k=5$ & sent addresses & $(S_k,L_k) = (8,24)$ & fee-reward ratio & $(k,z_u,z_l) = (5,1.5,0.5)$  \\[0.5ex] 
\midrule  
 2021  & confirmation time & $k=126$ & confirmation time & $(S_k,L_k) = (16,48)$ & sent USD & $(k,z_u,z_l) = (21,1.75,0.75)$  \\[0.5ex] 
- 2022& sent addresses & $k=21$ & sent addresses & $(S_k,L_k) = (4,12)$ & fee-reward ratio & $(k,z_u,z_l) = (5,1.5,0.5)$  \\[0.5ex]  
\midrule
2022 & sent addresses & $k=21$ & sent addresses & $(S_k,L_k) = (4,12)$ & sent USD & $(k,z_u,z_l) = (21,1.75,0.75)$  \\[0.5ex] 
- 2023 & coin days destroyed & $k=126$ & transactions & $(S_k,L_k) = (8,24)$ & coin days destroyed & $(k,z_u,z_l) = (5,1.75,0.75)$  \\[0.5ex] 
\bottomrule 
\end{tabular}
\label{table:StrategySelection}
\end{table*}

\begin{acks}
We would like to thank the Oxford-Man Institute of Quantitative Finance for computing support. 
\end{acks}

\printbibliography

\appendix
\newpage
\section{Feature-parameter Selections}
\label{apdx:FeatureParameterCombinations}
In Table \ref{table:StrategySelection}, we record the feature-parameter combinations selected by realistic traditional strategies in Section \ref{section:Results_RealisticStrategies}. BAZ performs best and correctly selects \textit{sent addresses} with short timescales in all test sets. In contrast, MOP only identifies the optimal timescale for \textit{sent addresses} in the final two test sets. REV consistently chooses \textit{sent USD}, which fails to generalise to unseen regimes: a positive $Sharpe = \textbf{1.26}$ in the first 3 test sets is offset by $Sharpe = \textbf{-2.31}$ in the final test set.

\section{MFIN Complexity}
\label{apdx:Complexity}

For a given number of assets $N_A$, the complexity of MFIN models depends on hyperparameters $N_H$, $n\_filters$, and $ts\_filter\_length$. To inspect the impact of one hyperparameter, we keep the others fixed at a median value. In OrigCIM-LSTM, the Feature Extractor contributes the majority of parameters (sometimes more than 90\%) via convolutional filters. With alternative DIN variants, such as DeepLOB-TFT, which only learn cross-factor interactions in the Position Sizer, there can be a more balanced contribution.


\begin{table}[!ht]
\centering
\caption{\centering{Number of trainable parameters $N_P$ for MFIN.}}
\begin{tabular}{ c c | c c c c}
\toprule
  & & \multicolumn{4}{c}{$N_A$} \\[0.5ex] 
  & & 1 & 7 & 20 & 50  \\[0.5ex] 
\midrule \midrule
 $N_H$ & 32 &  64K & 170K & 399K & 929K  \\[0.5ex] 
  & 64 &  80K & 185K & 415K & 945K   \\[0.5ex] 
 $(n\_filters = 40)$ & 96 &  103K & 209K & 438K & 968K \\[0.5ex] 
 $(ts\_filter\_length = 10)$ & 128 &  135K & 241K & 470K & 1,000K  \\[0.5ex] 
\midrule  
 $n\_filters$ & 16 &  49K & 66K & 103K & 190K  \\[0.5ex] 
  & 32 &  74K & 142K & 289K & 629K   \\[0.5ex] 
  $(N_H = 80)$ & 48 &  110K & 262K & 591K & 1,354K  \\[0.5ex] 
  $(ts\_filter\_length = 10)$ & 64 &  156K & 426K & 1,012K & 2,366K   \\[0.5ex] 
\midrule
  $ts\_filter\_length$ & 3 &  68K & 107K & 190K & 384K  \\[0.5ex] 
  & 5 &  74K & 132K & 257K & 547K   \\[0.5ex]
  & 10 &  90K & 196K & 425K & 955K   \\[0.5ex] 
  $(N_H = 80)$ & 15 &  106K & 260K & 593K & 1,363K  \\[0.5ex] 
  $(n\_filters = 40)$ & 20 &  122K & 324K & 761K & 1,771K   \\[0.5ex] 
\midrule \midrule
 \multicolumn{2}{c}{Datapoints} &  42K & 295K & 843K & 2,108K \\[0.5ex] 
\bottomrule 
\end{tabular}
\label{table:param_count}
\end{table}



\end{document}